\documentclass[%
superscriptaddress,
preprint,
 amsmath,amssymb,
 aps,
]{revtex4-2}
\usepackage[caption=false]{subfig}
\usepackage[normalem]{ulem} 
\usepackage[utf8]{inputenc}
\usepackage{textcomp}
\usepackage{xcolor}
\usepackage{graphicx}
\usepackage{amsmath}
\usepackage[version=4]{mhchem}
\usepackage{siunitx}
\usepackage{longtable,tabularx}
\usepackage{comment}
\usepackage{amsfonts}
\usepackage{graphics}
\usepackage{cleveref}
\usepackage{tikz}
\usepackage[outdir=./]{epstopdf}
\usepackage{import}
\newcommand{\R}{\mathbb{R}}
\newcommand{\msh}[1]{{\color{black}{#1}}}
\newcommand{\ak}[1]{{\color{black}{#1}}}

\begin{document}
\title{Nonlinear Stability Analysis of Transitional Flows using Quadratic Constraints} 
\author{Aniketh Kalur\footnote{Graduate Student, Aerospace Engineering and Mechanics}}
\affiliation{University of Minnesota, Minneapolis, MN, 55455}
\author{Peter Seiler\footnote{Associate Professor, Electrical Engineering and Computer Science}}
\affiliation{University of Michigan, Ann Arbor, MI, 48109}
\author{Maziar S. Hemati\footnote{Assistant Professor, Aerospace Engineering and Mechanics }} 
\affiliation{University of Minnesota, Minneapolis, MN, 55455}
\begin{abstract}
The dynamics of transitional flows are governed by an interplay between the non-normal linear dynamics and quadratic nonlinearity in the Navier-Stokes equations.
In this work, we propose a framework for nonlinear stability analysis that exploits the fact that nonlinear flow interactions are constrained by the physics encoded in the nonlinearity.
In particular, we show that nonlinear stability analysis problems can be posed as convex optimization problems based on Lyapunov matrix inequalities and a set of quadratic constraints that represent the nonlinear flow physics.
The proposed framework can be used to conduct global and local stability analysis as well as transient energy growth analysis.
The approach is demonstrated on the low-dimensional Waleffe-Kim-Hamilton model of transition and sustained turbulence.
Our analysis correctly determines the critical Reynolds number for global instability.
We further show that the lossless (energy conservation) property of the nonlinearity is destabilizing and serves to increase transient energy growth.
Finally, we show that careful analysis of the multipliers used to enforce the quadratic constraints can be used to extract dominant nonlinear flow interactions that drive the dynamics and associated instabilities.
\end{abstract}

\maketitle

\section{Introduction}
Many complex flow phenomena arise from the interplay 
between the non-normal linear dynamics and quadratic 
nonlinearity in the Navier-Stokes equations~(NSE).
%
%
In wall-bounded shear flows,
the high-degree of non-normality of the linearized NSE results in a transient energy growth~(TEG) of small flow perturbations~\citep{schmid2001,schmid2007,Trefethen1993}, even when the dynamics are linearly asymptotically stable.
%
%
%
As a result, linear stability analysis tends to over predict the critical Reynolds number~($Re_c$) for instability in many shear flows~\citep{schmid2007,schmid2001,ReddyJFM1993}.
The fact that the flow transitions below the predicted $Re_c$ is partly attributed to the non-modal growth that pushes the flow state away from the equilibrium base flow~\cite{Trefethen1993,schmid2001,schmid2007,ReddyJFM1993}.
%
%
%
Indeed, TEG is a necessary condition for transition~\cite{WaleffeSIAM1995,WaleffePOF1995}.
Nevertheless, non-modal TEG alone is not 
sufficient to cause transition:
%
it is the interaction of non-modal TEG with the nonlinearity that 
triggers secondary instabilities and drives 
the state outside the basin of attraction.
Without the nonlinear terms, the notion of a finite basin of attraction would not make sense.
Interestingly, although the nonlinearity is lossless and energy-conserving~\citep{Sharma2011,WaleffeSIAM1995}, it interacts with the linear dynamics in such a way as to increase the maximum transient energy growth~(MTEG) that can be realized~\cite{Kerswell2018}.
These transition scenarios cannot be fully analyzed without accounting for the nonlinear terms in NSE.

Analysis methods have been proposed to account for the interplay between the linear and nonlinear
terms in transitional and turbulent flows. 
One such approach is the resolvent analysis framework~\citep{McKeonJFM2010,JovanovicJFM2005,JovanovicAnnRev2020}, which leverages the fact that the NSE can be expressed as a feedback interconnection between a linear operator and a nonlinear operator---a so-called \emph{Lur'e decomposition}~\cite{Khalil}.
Resolvent analysis goes a step further to consider the nonlinearity 
as an implicit forcing input on the linear dynamics~\citep{McKeonJFM2010,TairaModal2017}.
This perspective greatly simplifies the resulting analysis problem, as only the linear system---described by the input-output properties of the 
linear resolvent operator---needs 
to be analyzed.
Within the context of turbulent flows, resolvent analysis provides information on how fluctuations in a time-averaged flow are attenuated or amplified from nonlinear effects.
Resolvent analysis has been successfully employed in the study of various flows~\cite{Taira2019,McKeon2017}, including pipe flows~\citep{SharmaJFM2013}, open cavity flows~\citep{Sun2019}, and flows over riblets~\citep{Chavarin2019}.

Related methods have been proposed to account for the nonlinearity more directly.
%
The passivity framework has been shown to be effective in flow control 
based on the nonlinear NSE~\citep{Heins2016,Damaren2016,Damaren2018}. 
In these studies, the passivity property~\citep{Khalil} of the nonlinear terms in the NSE are leveraged to design a linear controller that can stabilize the system. 
%
Further advances have been made in input-output methods to study performance, worst-case amplification, stability, and transition for NSE using dissipation inequalities~\citep{Ahmadi2018}. 
%
Dissipation inequalities derived from NSE can be posed as linear matrix inequality~(LMI) problems, which are then solved using convex optimization methods to analyze various wall-bounded shear flows. 
These techniques generalize the classical energy-based analysis approaches~\cite{schmid2001,josephBook} and also have close ties with nonlinear Lyapunov stability analysis approaches developed for NSE based on sum-of-squares~(SOS) optimization~\cite{goulart2012}.




In this paper,
we propose an alternate framework for nonlinear stability analysis
that uses quadratic constraints to  account for nonlinear 
flow interactions with minimal complexity.  
The approach is predicated on the fact that nonlinear flow interactions are 
constrained by the physics encoded within the nonlinear terms in the NSE---e.g.,~the nonlinearity is quadratic, energy conserving, and lossless.
Mathematically, these physics can be expressed as quadratic constraints between the 
inputs and outputs of the nonlinearity.
In turn, these quadratic constraints serve as reduced-complexity models for the nonlinear terms, and can be incorporated within a Lyapunov-based analysis to  perform reliable stability and input-output analysis in the nonlinear setting.
The general framework introduced here is applicable to any system that has (non-normal) linear dynamics acting in feedback with a lossless nonlinearity---the NSE being a special case.
To establish a proof-of-concept, we formulate and demonstrate the proposed analysis framework on the nonlinear Waleffe-Kim-Hamilton (WKH) model of transition and sustained turbulence~\citep{WaleffeSIAM1995}. 
As with the NSE, the WKH model admits a Lur'e decomposition with non-normal linear dynamics and a quadratic lossless nonlinearity, making it relevant for 
formulating and demonstrating the proposed quadratic constraints framework for 
nonlinear stability analysis of fluid flows.


The paper proceeds as follows. In Section~\ref{sec:WKH}, we introduce the WKH model in Lur'e form. We then introduce the quadratic constraints framework and associated stability analysis problem in Section~\ref{sec:QC_stability}.  In Section~\ref{sec:QC_lossless}, we account for the energy conserving nonlinearity in global stability analysis via the addition of a quadratic lossless constraint. In Section~\ref{sec:QC_local}, we show that additional quadratic constraints can be introduced to conduct local stability analysis, which is needed when $Re>Re_c$ and the equilibrium point is no longer globally asymptotically stable. In Section~\ref{sec:QC_MTEG}, we formulate an analysis problem to determine the maximum transient energy growth~(MTEG) that can be realized by the system dynamics. In Section~\ref{sec:Lagrange_Mul_Analysis}, we show that we can obtain insights into dominating nonlinear flow interactions that underlie the dynamics by analyzing the multipliers used to enforce the constraints within the analysis framework. Finally, we provide concluding remarks of our study in Section~\ref{sec:Conclusion}.

The section-wise specific contributions of this paper are as follows:

\begin{enumerate}
    
    \item Section~\ref{sec:QC_lossless}: We find that the lossless constraint alone enables prediction of the critical Reynolds number ($Re_c$) for global instability, consistent with the $Re_c$ found by other means in~\citep{WaleffePOF1995}.
    
    \item Section~\ref{sec:QC_local} and~\ref{sec:local_stab_analysis}: The ``local'' quadratic constraints represent the influence of the nonlinearity when the flow is restricted to a local neighborhood about the equilibrium point. By modifying the size of the local region, the influence of the nonlinearity can be assessed and the nonlinearity is found to have a destabilizing effect. 
    
    \item Section~\ref{sec:QC_MTEG}: We show that the nonlinearity contributes to increasing the MTEG relative to a purely linearly analysis, consistent with findings from nonlinear non-modal stability analysis of NSE~\cite{Kerswell2018}. Indeed, these findings confirm previous findings that the optimal disturbance for the nonlinear system is different from the linear optimal disturbance that is often used to study transition~\cite{Kerswell2018}.
    
    \item Section~\ref{sec:Lagrange_Mul_Analysis}: We demonstrate that---without any a priori knowledge---the proposed analysis approach is able to extract the same dominant nonlinear flow interactions whose physical importance is argued in~\citep{WaleffePOF1995,WaleffeSIAM1995}.

\end{enumerate}

\section{Waleffe-Kim-Hamilton Model of Transition and Sustained Turbulence}
\label{sec:WKH}	


	The Waleffe-Kim-Hamilton~(WKH) model is a low-order mechanistic model for transition and sustained turbulence in shear flows.  The model is based on observations from direct numerical simulations of a plane Couette flow~\citep{WKHOrigins}, and was introduced to highlight the importance of nonlinear interactions with the non-normal linear dynamics in the NSE. The WKH model was studied in greater detail by Waleffe in~\citep{WaleffePOF1995} and is given by,

    \begin{align}
        \begin{bmatrix} \dot{u} \\
        \dot{v}\\
        \dot{w}\\
        \dot{m}\end{bmatrix} &= \frac{1}{Re}\begin{bmatrix} 0\\
        0\\
        0\\
        \sigma
        \end{bmatrix} -\frac{1}{Re}\begin{bmatrix} \lambda u\\
        \mu v\\
        \nu w\\
        \sigma m
        \end{bmatrix} + \begin{bmatrix} 
        0 & 0 & -\gamma w & v\\
        0 & 0 & \delta w & 0\\
        \gamma w& -\delta w & 0 & 0\\
        -v & 0 & 0 & 0
        \end{bmatrix} \begin{bmatrix}
        u\\
        v\\
        w\\
        m\end{bmatrix}.
        \label{eq:orig_wkh}
        \end{align}
Here, $Re$ denotes the Reynolds number; $u$ represents the amplitude of the spanwise modulation of streamwise
	velocity; $v$ represents the amplitude of the streamwise rolls; $w$ represents the
	amplitude of the inflectional streak instability; and $m$ represents the amplitude
	of the mean shear~\citep{WaleffePOF1995}. The constants $\lambda$, $\mu$, $\nu$, $\sigma$ are positive parameters corresponding to viscous decay rates. The constants $\gamma$ and $\delta$ represent nonlinear interaction
	coefficients and should have the same sign~\citep{WaleffePOF1995}.
		
	The model admits a laminar equilibrium point at $(u, v, w, m)_e =
	(0, 0, 0, 1)$. 
	For the proposed stability analysis, we perform a change of coordinates to translate the equilibrium point of Eq.~\eqref{eq:orig_wkh} to the origin. The equilibrium point in these new coordinates is $\mathbf{x}_{e} = (0,0,0,0)$ and the state is $\mathbf{x}= (u,v,w,\bar{m})$, where $\bar{m} = m-1$. The system in this translated coordinate system is,

\begin{align}
\underbrace{\begin{bmatrix}
    \dot{u}\\
    \dot{v}\\
    \dot{w}\\
    \dot{\bar{m}}
    \end{bmatrix}}_{\dot{\mathbf{x}}} &= \underbrace{\begin{bmatrix} -\frac{\lambda}{Re} & 1 & & \\
                                    & -\frac{\mu}{Re} & & \\
                                    & & -\frac{\nu}{Re} &\\
                                    & & & -\frac{\sigma}{Re}\end{bmatrix}\begin{bmatrix} u\\
                                    v\\
                                    w\\                                    
                                    \bar{m}\end{bmatrix}}_{A\mathbf{x}} + \underbrace{\begin{bmatrix} 0 & 0 & -\gamma w & v\\
                                    0 & 0 & \delta w & 0\\
                                    \gamma w & -\delta w & 0 & 0\\
                                    -v & 0 & 0 & 0 \end{bmatrix}\begin{bmatrix} u\\
                                    v\\
                                    w\\
                                    \bar{m}\end{bmatrix}}_{N(\mathbf{x})=Q(\mathbf{x})\mathbf{x}},
                                    	    \label{eq:xform_wkh}
\end{align}
which makes the non-normality of the linear dynamics explicit~\citep{HenningsonPOFComment}.  
    
\begin{figure}[!htb]
    \centering
    \includegraphics[scale=0.5]{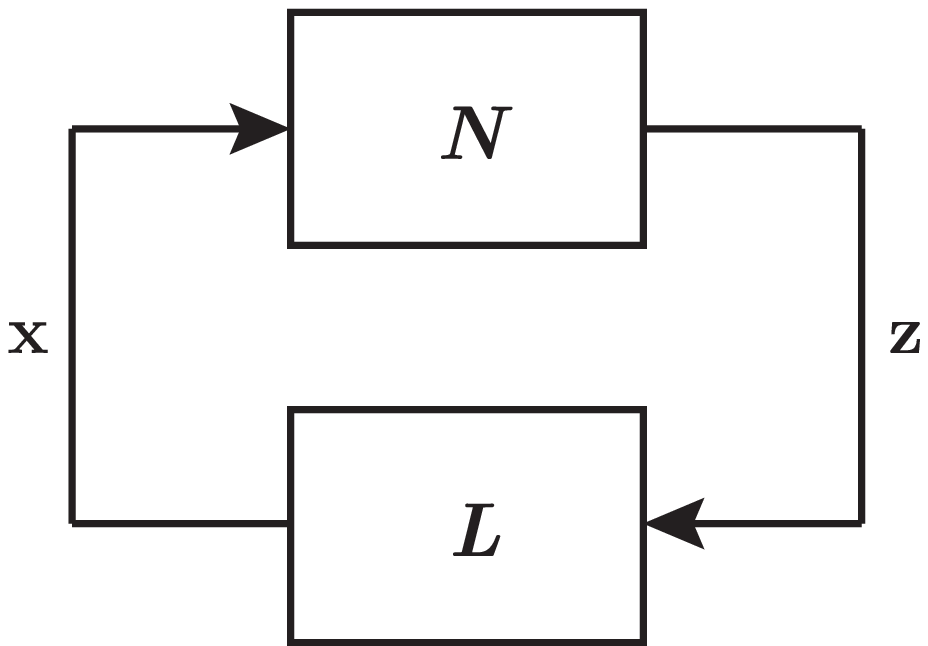}
    \caption{Lur'e representation of the  WKH system $F_u(L,N).$}
    \label{fig:Lure_WKH}
\end{figure}

The WKH system in Eq.~\eqref{eq:xform_wkh} can be represented as

\begin{align}
     \dot{\mathbf{x}} &= A \mathbf{x} + N(\mathbf{x}),
     \label{eq:wkh_oneline}
\end{align}
where the linear operator $A$ is non-normal and asymptotically stable, and $N(\mathbf{x})$ is a quadratic nonlinearity given by $N(\mathbf{x}) = Q(\mathbf{x})\mathbf{x}$.  Note that the nonlinear term is skew-symmetric: i.e., $Q(\mathbf{x})  = -Q(\mathbf{x})^T\in \mathbb{R}^{4 \times 4}$. The linear and nonlinear terms can be partitioned into Lur'e form~\citep{Khalil}, with the two systems acting in feedback with each other~(see FIG.~\ref{fig:Lure_WKH}):
\begin{subequations}
\begin{align}
    \dot{\mathbf{x}} &= L(\mathbf{x},\mathbf{z}) := A\mathbf{x} + \mathbf{z} \\
    \mathbf{z} &= N(\mathbf{x}) 
    \label{eq:z}
\end{align}
\label{eq:wkh_ss}
\end{subequations}
where $\mathbf{z}\in \mathbb{R}^4$.
This Lur'e decomposition of the WKH system is denoted as $F_u(L,N)$. 

In all that follows, the values for the decay rates are chosen to be $\lambda = \mu =\sigma = 10 $ and $\nu =15$. The nonlinear interaction coefficients are selected to be $\gamma=0.1$ and $\delta = 1$. These values are chosen based on the system studied in~\citep{WaleffePOF1995}. In the remainder of this work, the only parameter that is varied for stability and transient energy growth analyses is $Re$.

\section{Nonlinear Stability Analysis using Quadratic Constraints}
\label{sec:QC_stability}

Lyapunov stability methods~\citep{Khalil} can be used to analyze the stability of a system given by Eq.~\eqref{eq:wkh_ss}. Here, the stability is analyzed around the equilibrium point $\mathbf{x}_e=\mathbf{0}$. To analyze stability using Lyapunov stability methods, we define a quadratic scalar energy function $V: \mathbb{R}^n \rightarrow \mathbb{R}$. The energy function $V(\mathbf{x}) = \mathbf{x}^TP\mathbf{x}$ is a candidate Lyapunov function~\citep{Khalil}. From the Lyapunov stability theorem, the equilibrium point $\mathbf{x}_e=\mathbf{0}$ is globally asymptotically stable when $dV(\mathbf{x})/dt <0~\forall~\mathbf{x}\neq \mathbf{0}$ and $P>0$. In other words,  the system is globally asymptotically stable around the equilibrium point $\mathbf{x}_e=\mathbf{0}$ if the energy continuously decreases in time. The time derivative of the Lyapunov function for the nonlinear system in Eq.~\eqref{eq:wkh_ss}, along trajectories of the system is given by:
\begin{align}
\begin{split}
\label{eq:Vdot}
\frac{d}{dt} V(\mathbf{x}) &= 2 \mathbf{x}^T P ( A\mathbf{x} + \mathbf{z} ) \\
                     &= 2 \mathbf{x}^T  P ( A\mathbf{x} + N(\mathbf{x}) ).
                     \end{split}
\end{align}

Including the effects of N($\mathbf{x}$) to analyze stability is crucial to understanding the global asymptotic stability of the nonlinear system.  However, accounting for the nonlinear term N($\mathbf{x}$) complicates the stability analysis and a quadratic Lyapunov function will not necessarily be a good choice as a candidate Lyapunov function. Here, we leverage the fact that the inputs and outputs of the nonlinearity $N$ satisfy a set of  quadratic constraints, 
thereby enabling stability analysis of the whole feedback interconnection $F_u(L,N)$ without the full complexity involved in an explicit treatment of the nonlinearity.

To do so, we first show that stability analysis benefits from consideration of the nonlinear term as energy conserving and lossless, neither producing nor dissipating energy. The lossless property can be represented as a quadratic constraint to represent the nonlinear term within the Lyapunov analysis. The lossless constraint captures global behavior of the nonlinearity. Further, we analyze local behavior of the nonlinearity around a neighborhood by representing its local properties as ``local'' constraints. We also show that local nonlinear properties  play a role in destabilizing the system, whereas a linear stability analysis predicts the WKH system to be globally asymptotically stable for all $Re$. Both global and local stability analysis and the results are discussed in the following sections. 


\subsection{Stability: Representing Lossless Nonlinearity with Quadratic Constraints}
\label{sec:QC_lossless}
The nonlinear term in Eq.~\eqref{eq:xform_wkh} is skew-symmetric, therefore
\msh{\begin{align}
\mathbf{x}^T N(\mathbf{x}) = \mathbf{x}^T Q(\mathbf{x}) \mathbf{x} = 0,~\forall~ \mathbf{x}.
    \label{eq:global_property}
\end{align}}
\msh{The physical interpretation of this property is that the nonlinearity is energy conserving, serving only to redistribute energy between modes.}
This ``lossless'' property of the nonlinear term is also observed in many wall-bounded shear flows~\citep{Sharma2011}. 
\msh{The stability analysis reduces to the following question: Does the constraint in Eq.~\eqref{eq:global_property} imply  $\dot{V}(\mathbf{x})<0$ in Eq.~\eqref{eq:Vdot} for all $\mathbf{x} \ne \mathbf{0}$? The answer is yes, if there exists a $P>0$ and a Lagrange multiplier $\xi_{p_0}$ (positive or negative) such that



\begin{align}
2 \mathbf{x}^T  P ( A\mathbf{x} + N(\mathbf{x})) + 2\xi_{po} \mathbf{x}^TN(\mathbf{x}) <0,
\label{eq:vdot_infeasible}
\end{align}
which essentially says that the energy function $V(\mathbf{x})$ decreases for any $\mathbf{x}$ and $N(\mathbf{x})$ satisfying the lossless constraint in Eq.~\eqref{eq:global_property}. 
%

Consider now that the lossless property in Eq.~\eqref{eq:global_property} can be 
expressed equivalently as a quadratic constraint between 
the inputs $\mathbf{x}$ and outputs $\mathbf{z}=N(\mathbf{x})$ of the nonlinearity:}
\begin{align}
    \centering
    \begin{pmatrix} \mathbf{x}\\ \mathbf{z}\end{pmatrix}^T \underbrace{\begin{pmatrix} \mathbf{0} & \mathbf{I}\\
    \mathbf{I} & \mathbf{0} \end{pmatrix}}_{:= M_0}\begin{pmatrix} \mathbf{x}\\ \mathbf{z}\end{pmatrix} = 0,~\forall~\mathbf{x}~\text{and}~\mathbf{z} \in \mathbb{R}^4,
    \label{eq:lossless_constraint}
\end{align}
where $\mathbf{0}, \mathbf{I} \in \R^{4\times 4}$ denote the zero and identity matrices, respectively. 
\msh{
Thus, Eq.~\eqref{eq:vdot_infeasible} can be recast as,
\begin{align}
 \begin{bmatrix} \mathbf{x}\\\mathbf{z} \end{bmatrix}^T \left\{ \begin{bmatrix} A^TP+PA & P\\ P & \mathbf{0} \end{bmatrix}  + \xi_{p_0}M_0 \right \}\begin{bmatrix} \mathbf{x}\\\mathbf{z} \end{bmatrix}< 0.
 \label{eq:stability_lmi_infeasible}
\end{align}
The lossless constraint $\mathbf{z}^T\mathbf{x}=0$ is captured by the block matrix $M_0 \in \R^{8 \times 8}$ defined in Eq.~\eqref{eq:lossless_constraint}.
We note that the strict inequality in Eq.~\eqref{eq:vdot_infeasible} cannot be satisfied because the sum in brackets yields $\mathbf{0}$ for the principal sub-matrix in the lower-right block.
As such, we introduce a positive perturbation on Eq.~\eqref{eq:vdot_infeasible} to relax the requirement for a strict inequality as,
\begin{align}
2 \mathbf{x}^T  P ( A\mathbf{x} + N(\mathbf{x})) + 2\xi_{po} \mathbf{x}^TN(\mathbf{x}) +2\epsilon\mathbf{x}^TP\mathbf{x}\le0,
 \label{eq:Vdot_constraints}
\end{align}
where $\epsilon>0$. 
This condition is equivalent to $\dot{V}(\mathbf{x})\le-\epsilon V(\mathbf{x})$ for all $\mathbf{x}\ne\mathbf{0}$, which guarantees exponential stability with a minimum convergence rate of $\epsilon$.

The stability condition in Eq.~\eqref{eq:Vdot_constraints} can be recast in terms of the quadratic lossless constraint in Eq.~\eqref{eq:lossless_constraint} to yield,
\begin{align}
 \begin{bmatrix} \mathbf{x}\\\mathbf{z} \end{bmatrix}^T \left\{ \begin{bmatrix} A^TP+PA & P\\ P & \mathbf{0} \end{bmatrix}  + \xi_{p_0}M_0 + \begin{bmatrix} \epsilon P & \mathbf{0} \\
    \mathbf{0} & \mathbf{0} \end{bmatrix}  \right \}\begin{bmatrix} \mathbf{x}\\\mathbf{z} \end{bmatrix}\le 0.
 \label{eq:stability_lmi}
\end{align}
%
Unlike the stability condition in Eq.~\eqref{eq:stability_lmi_infeasible}, it is possible for this new stability condition in Eq.~\eqref{eq:stability_lmi} to be satisfied.
It is interesting to note that a feasible solution to Eq.~\eqref{eq:stability_lmi}  must satisfy $P=-\xi_{p0}\mathbf{I}$ for some $\xi_{p0}<0$.
This condition further implies that feasibility of the stability condition in Eq.~\eqref{eq:stability_lmi} requires $A+A^T+\epsilon\mathbf{I}<0$.
In the limit $\epsilon\rightarrow0$, this condition is equivalent to $A+A^T<0$, which is necessary and sufficient condition for unity maximum transient energy growth due to linear non-modal dynamics~\cite{Whidborne2007}.
%
}

\msh{In light of the stability  condition in \eqref{eq:stability_lmi}, it follows that stability of the linear element $L$ and a lossless nonlinearity can be formulated as an LMI feasibility problem in the variables $P>0$ and $\xi_{p0}$.
In particular, the system $F_u(L,N)$ is globally asymptotically stable if there exists $P>0$ and $\xi_{p_0}$ such that the following LMI holds for a given $\epsilon>0$:}
\begin{align}
    \begin{bmatrix} A^TP+PA & P\\ P & \mathbf{0} \end{bmatrix}  + \xi_{p_0}M_0   +\begin{bmatrix} \epsilon P & \mathbf{0}\\
    \mathbf{0} & \mathbf{0} \end{bmatrix}\le 0.
    \label{eq:global_stab_thm}
\end{align}
 The feasibility of the LMI in Eq.~\eqref{eq:global_stab_thm} is only sufficient to establish the global asymptotic stability of the WKH system, as it only relies on the lossless property and does not depend on any other specific details of the nonlinearity. The LMI with constraints in Eq.~\eqref{eq:global_stab_thm} is a convex optimization problem that can be solved using standard numerical tools. In the remainder of this work, we use CVX~\citep{cvx,gb08}, which is a package for specifying convex optimizations, combined with the commercially available solver MOSEK~\citep{mosek}.

To analyze the global stability of the WKH system, we solve the LMI in Eq.~\eqref{eq:global_stab_thm} \msh{with $\epsilon=10^{-6}$} for variables $P$ and $\xi_{p_0}$ at different values of $Re$. On performing the global stability analysis using the lossless constraint, we find that the WKH model for the  given parameter values is globally asymptotically stable for \ak{ $Re<20$.}
This finding is consistent with $Re_c=20$ for global asymptotic stability reported by Waleffe~\citep{WaleffePOF1995}.


The nonlinear system's global asymptotic stability regime can be determined using the lossless constraint alone. Therefore, the lossless constraint is sufficient for characterizing the global stability of the nonlinear WKH system. Note that the linear WKH system is globally asymptotically stable for all $Re$, and so the nonlinear term is destabilizing. In the nonlinear WKH system there exist certain initial conditions for which the system trajectories will fail to converge to the equilibrium for \msh{$Re\ge20$}.
To investigate this further, we propose a set of local constraints on the nonlinearity that enable a local stability analysis, as described in the next section.

\subsection{Stability: Representing Local Properties of Nonlinearity using Quadratic Constraints}\label{sec:QC_local}
{\color{black}
\ak{ The WKH system with the lossless constraint is globally asymptotically stable for $Re\le 20$ for the W parameters and for $Re\le 2$ for the B\&T parameters. To analyze the system for larger $Re$,  we propose a ``local'' stability analysis
as follows:}  Select a local
neighborhood \ak{$\|\mathbf{x}\|^2\le R^2$} around the equilibrium point $\mathbf{x}_e=\mathbf{0}$. Local analysis restricts the state $\mathbf{x}$ to lie in a local region $R$, which result in ``local'' constraints
for $N(\mathbf{x})$ within this local region.  The analysis condition, given below,
attempts to use these local quadratic constraints to show that: (i)~the system state 
remains within the  local region and (ii)~it converges asymptotically back to $\mathbf{x}_e=\mathbf{0}$. These quadratic constraints are tighter (more powerful) for smaller values of $R$ and become 
looser (less powerful) as $R$ becomes larger.  Thus, these local analysis results provide
a range of results between global asymptotic stability (roughly as $R\rightarrow\infty$) and 
stability of the linearized system (roughly as $R\rightarrow 0$). 
\ak{We will show later in this section that $R$ can be used to estimate the region of attraction~(ROA) for the equilibrium point.}


Recall that the nonlinearity in the WKH model is quadratic and can be expressed as  $\mathbf{z} = \mathbf{x}^TQ(\mathbf{x})\mathbf{x}$~(see Eq.~\eqref{eq:xform_wkh}). To illustrate the approach, first consider the  scalar example  $z = x^2$ (green curve in FIG.~\ref{fig:quadratic_function}). Within a given region \ak{$| \mathbf{x}|<R$}, the output satisfies \ak{$z^2 = x^4 < R^2 x^2$}. Which further implies that \ak{$|z|< R|x|$}, where $R$ is the slope of the line. The quadratic function is restricted by the bound R, but this bound would graphically correspond to drawing a line of slope $+R$ and $-R$ (red lines in FIG.~\ref{fig:quadratic_function}). The slope $R$ can have a large value or a small value, as illustrated in FIGs.~\ref{fig:largeR} and \ref{fig:smallR}, respectively. If $x$ remains in the interval $[-R,+R]$, then the nonlinear function lies between these two linear lines with slope $\pm R$~(gray shaded region in FIGs.~\ref{fig:largeR} and \ref{fig:smallR}). The dashed blue line in both these figures represents the maximum possible value of the pair ($x,z$), such that \ak{$|x|\le R$} for a given slope. It can be seen that as the slope $R$ is made larger~(FIG.~\ref{fig:largeR}), then the pair ($x,z$) also gets bigger, thereby moving the blue dashed line further away from the origin. Similarly, as slope of $R$ is made smaller~(FIG.~\ref{fig:smallR}), the pair ($x,z$) gets smaller, thereby moving the dashed line towards the origin, which corresponds to a reduction in maximum value of $z$. Finally, note that as the slope $R$ tends to zero, the sector shrinks to zero. Thus, $R\rightarrow 0$ corresponds to a nonlinear term with zero output---equivalent to a linear analysis.  Conversely, as $R\rightarrow \infty$, then this sector becomes arbitrarily large and provides essentially no information---corresponding to a global analysis.

The sector formed by lines of slope $\pm R$ facilitates bounding the pair $(x,z)$ to perform analysis in a localized setting, where the value of $R$ also determines the amount of nonlinear behavior captured by the local constraint. A brief introduction to scalar sector bounded nonlinearities is presented in Appendix~\ref{sec:appendixB}. The remainder of this section generalizes this basic concept to the multivariable quadratic terms that appear in the WKH model.

\begin{figure}
    \centering
    \subfloat[Large slope $R$ forms larger sectors]{\includegraphics[scale=0.5]{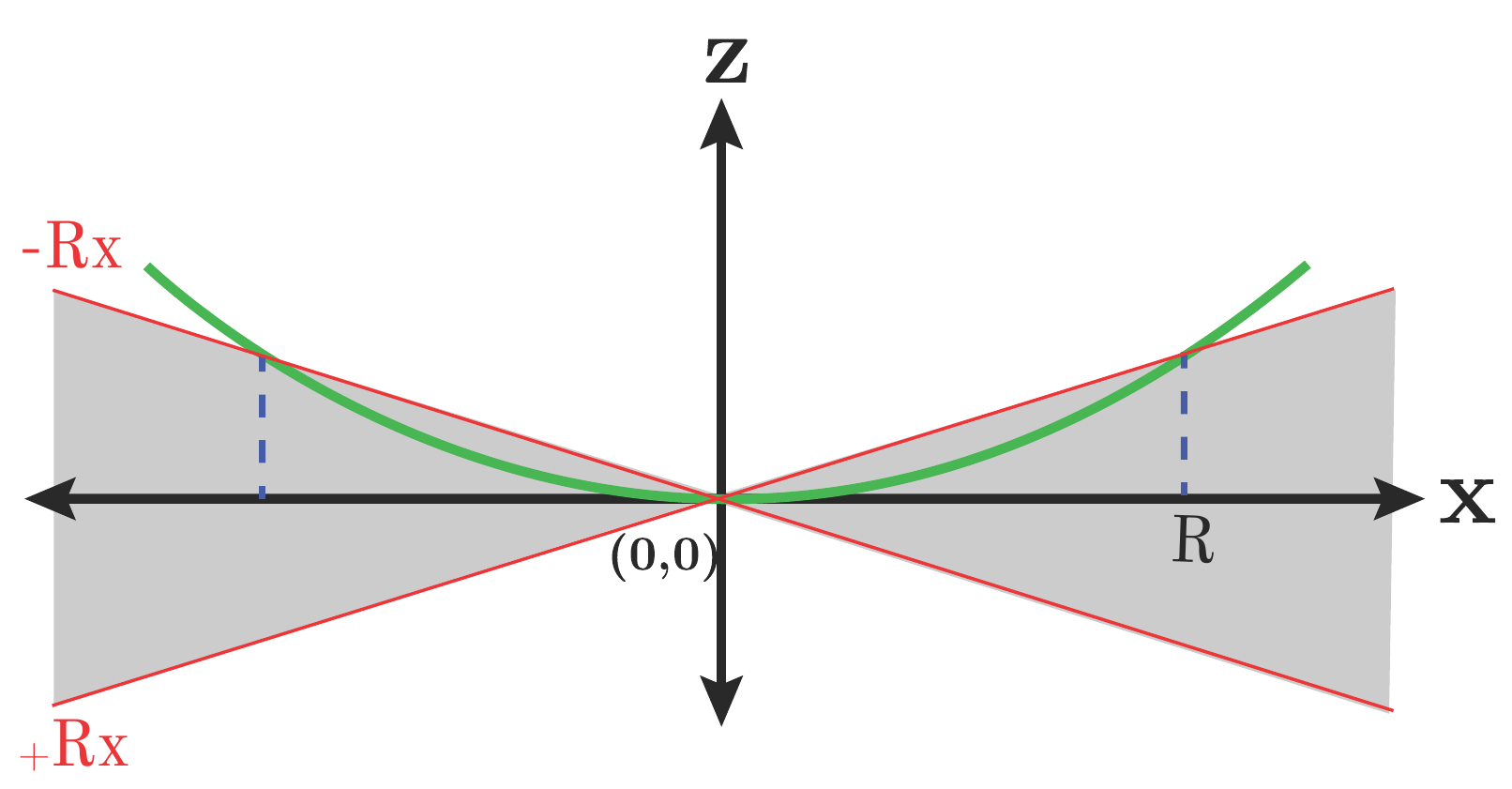}\label{fig:largeR}}
    \hfill
    \subfloat[Small slope $R$ forms smaller sectors]{\includegraphics[scale=0.5]{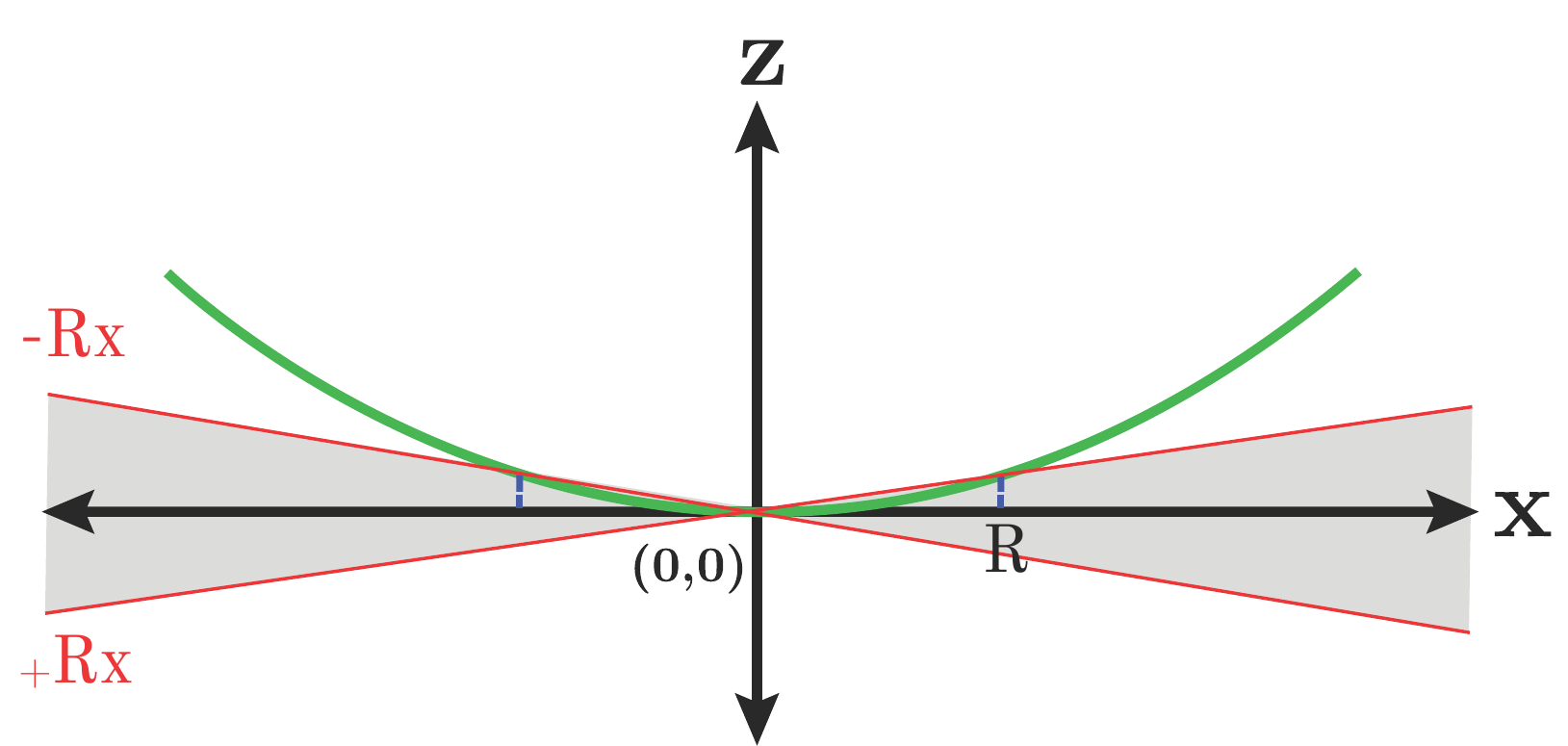}\label{fig:smallR}}
    \caption{Illustration of a scalar quadratic function $x=z^2$ that lies inside the sector formed by lines of slope $-R$ and $R$~(red).  The blue dashed lines mark the maximum value of the pair $(x,z)$ for a given slope such that $|x|\ak{\le} R$.}
    \label{fig:quadratic_function}
\end{figure}
}
 From Eq.~\eqref{eq:xform_wkh}, each individual nonlinear term can be expressed as a quadratic function:

\begin{align}
    \mathbf{z} = \begin{bmatrix} z_1\\
    z_2\\
    z_3\\
    z_4\end{bmatrix} = \begin{bmatrix} \mathbf{x}^TQ_1\mathbf{x} \\
    \mathbf{x}^TQ_2\mathbf{x} \\
    \mathbf{x}^TQ_3\mathbf{x} \\
    \mathbf{x}^TQ_4\mathbf{x} \\\end{bmatrix}.
    \label{eq:local_nonlinear_terms}
\end{align}

Here each $Q_i \in \R^{4 \times 4}$ is a symmetric matrix.  Hence each $Q_i$ has real eigenvalues, and the spectral radius $\rho(Q_i)$ denotes the largest (magnitude) of these eigenvalues~\citep{Horn1190}.  Moreover, quadratic terms with symmetric matrices are upper bounded as follows~\citep{Horn1190}:

\begin{align}
   \label{eq:specrad}
   |z_i| = |\mathbf{x}^T Q_i \mathbf{x}| \le \rho(Q_i) \mathbf{x}^T \mathbf{x},~\text{for}~i=1~\text{to}~4.
\end{align}
Next, assume the state $\mathbf{x}$ remains within a ball of radius $R$, i.e. $\mathbf{x}^T \mathbf{x} \le R^2$. We can then square Eq.~\eqref{eq:specrad} to obtain the following constraint:
\begin{align}
    z_i^2 \leq \underbrace{\rho(Q_i)^2R^2}_{\alpha_i(R)^2} \mathbf{x}^T\mathbf{x},~\text{for}~i=1~\text{to}~4.
    \label{eq:quad_const}
\end{align}
This is a constraint involving squares of $\mathbf{x}$ and $z_i$. It can be written in a more useful quadratic constraint form.  Let $E_i \in \R^{4 \times 4}$ denote the matrix with the diagonal $(i,i)$ entry equal to one and all other entries equal to zero.  The constraint in Eq.~\eqref{eq:quad_const} is equivalent to:
\begin{align}
    \begin{bmatrix} \mathbf{x}\\
    \mathbf{z} \end{bmatrix}^T \underbrace{\begin{bmatrix} \alpha_i(R)^2 \mathbf{I} & \mathbf{0} \\
    \mathbf{0} & -E_i\end{bmatrix}}_{M_i} \begin{bmatrix} \mathbf{x}\\
    \mathbf{z} \end{bmatrix} \geq 0,~\text{for}~i=1~\text{to}~4.
    \label{eq:const_LMI}
\end{align}
 The above multivariable quadratic constraint in Eq.~\eqref{eq:const_LMI} is similar to the sector constraint in the scalar case, shown in Eq.~\eqref{eq:appendix_scalar_sector_constraints} of Appendix~\ref{sec:appendixB}. The above constraint provides a bound on the nonlinear term $z_i$ that holds over the local region $\mathbf{x}^T\mathbf{x} \le R^2$.  A local bound can be obtained for each of the four quadratic nonlinearities in Eq.~\eqref{eq:const_LMI}.  
 \ak{It should be noted that the lower right block in the $M_i$ matrices is non-zero, and so we can use the strict inequality $\dot{V}(\mathbf{x})+\xi_{p0}M_0+\sum_{i=0}^{4}\xi_{pi}M_i<0$.}
\msh{We will make use of these local constraints to study local stability of the WKH system in section~\ref{sec:local_stab_analysis}, and show that they can be used for transient energy growth analysis as well in Section~\ref{sec:local_MTEG_Analysis}.}

\subsection{Results: Local stability analysis using quadratic constraints}
\label{sec:local_stab_analysis}
The lossless property in Eq.~\eqref{eq:lossless_constraint} captures the global behavior of the quadratic nonlinearity. Given that the WKH system is not globally stable for $Re >20$, it is still beneficial to understand its local stability properties.  The linearization around $\mathbf{x}_e=\mathbf{0}$ is stable for all $Re>0$ because $A$ is Hurwitz. A more quantitative local stability analysis can be performed around $\mathbf{x}_e=\mathbf{0}$ using the local constraints derived in Eq.~\eqref{eq:const_LMI}. For the given constraint, by selecting a bound on the state $\|\mathbf{x}\|^2\le R^2$, we can determine if the states will return back to the origin in this local neighborhood $R$. The local stability analysis for the nonlinear system is performed by solving the following LMI feasibility problem:
 \begin{align}
\begin{split}
& P \geq~\text{I} \\
&  \xi_{p_i} \geq 0 \qquad (\text{for}~i=1~\text{to}~4)\\
&  \begin{bmatrix} A^TP + PA & PB\\
B^TP & \mathbf{0}\end{bmatrix} + \xi_{p_0}M_0 + \sum_{i=1}^4 \xi_{p_i} M_i <0,
\end{split}
\label{eq:Local_Stab_LMI}
\end{align}
here $\xi_{p_i}$~($i=1~\text{to}~4$) are Lagrange multipliers for local constraints. These Lagrange multipliers also provide information on which constraints are most relevant for proving stability on the local region.

It can also be shown that the level set $V(\mathbf{x})$ contained within $\|\mathbf{x}\|^2<R^2$ is invariant. Therefore, the state  once inside this set will always remain inside the set. This argument can be formalized using Lyapunov theory as shown in Appendix~\ref{sec:appendixA}. Solving the local stability problem in Eq.~\eqref{eq:Local_Stab_LMI} ensures that the states will decay back to the equilibrium $\mathbf{x}_e=\mathbf{0}$ for the nonlinear system.

Towards performing local analysis, we determine the largest possible lower bound on local stability region $R$ for a given $Re$.  In this neighborhood of size $R$, all trajectories will decay back to $\mathbf{x}_e=\mathbf{0}$. The analysis condition in Eq.~\eqref{eq:Local_Stab_LMI} can be used to determine the local stability region $R$ as a function of $Re$. Note that the linear state matrix $A$ inversely depends on $Re$ and the constraint matrices $M_i$ depend on $R^2$. For a given $Re$, we perform a bisection to compute the smallest value of $R$ such that Eq.~\eqref{eq:Local_Stab_LMI} is feasible.
%
The resulting relationship between $Re$ and $R$ is shown in FIG.~\ref{fig:local_stability}.
Note that $R$ decreases monotonically as $Re$ tends to $\infty$. This implies that the local stability region shrinks as $Re$ increases. On the other hand, $R$ tends to $\infty$ as $Re$ decreases to 20. In this case, the local stability region is increasing in size. This is consistent with the previous global stability result for $Re\le 20$. Note that the Lagrange multipliers $\xi_{pi}$ tend to zero as $Re$ tends down to 20. This indicates that the local constraints provide no useful information over that already provided by the lossless constraint for this case. For small values of $R\approx10^{-3}$ and below, we find that the nonlinear analysis corresponds to the linear analysis.



\begin{figure}[!htb]
   \centering
  \includegraphics[scale=0.6,trim=0cm 1cm 1cm 0.5cm]{Figs_New_V1/ReVsRopt_v2.eps}
  \vspace{0.25cm}
  \caption{As the $Re$ is increased, the local stability region $R$ decreases. As we approach $Re=20$, the size of $R\rightarrow \infty$, since we are approaching global stability regimes. As $Re\rightarrow \infty$, the size of region $R\rightarrow 0$, which corresponds to the linear analysis of infinitesimal perturbations.}
    \label{fig:local_stability}
\end{figure} 

 

\section{Nonlinear Transient Energy Growth Analysis using Quadratic Constraints}
\label{sec:QC_MTEG}
For an asymptotically stable linear system $\dot{\mathbf{x}}= A\mathbf{x}$, the state trajectories $\mathbf{x}(t) \rightarrow \mathbf{0}$ for any initial condition. If the matrix $A$ is non-normal, then the system energy will grow on transient time scales before decaying back to zero. In general, transient energy usually requires appropriate weights ($W$) to weight states such that $E=\tilde{\mathbf{x}}^TW\tilde{\mathbf{x}}$; however, for the remaining results and without loss of generality, we perform a similarity transform such that the energy $E=\mathbf{x}^T\mathbf{x}$. The peak of this energy growth curve is called the maximum transient energy growth~(MTEG) and is defined as:
\begin{align}
  \text{MTEG} := \max_{t \ge 0} \max_{\|\mathbf{x}(0)\|=1} \| \mathbf{x}(t) \|^2 \,\, .
\end{align}
 The MTEG represented by the variable $q^*$ is also defined as follows $q^* := \lambda_{max}(P)  \lambda_{max} (P^{-1})$ such that $P=P^T>0$ and $P$ satisfies $AP+A^TP <0$. The minimal upper bound can be obtained by solving the following optimization problem~\citep{Whidborne2007,BoydLMI1994}

\begin{align}
\begin{split}
 q^*:=  \min  \quad & q\\
\mbox{ subject to:}~&  \text{I}\leq P \leq q\text{I}~, \\
&   A^TP + PA<0. \\
  \end{split}
  \label{eq:linear_MTEG}
\end{align}
This optimization has LMI constraints and a linear cost involving variables $(P,q)$. This optimization is known as a semidefinite program~(SDP).  The LMI constraints imply that $V(\mathbf{x}):= \mathbf{x}^T P \mathbf{x}$ is a Lyapunov function for the system such that $V(\mathbf{x}(t)) \le V(\mathbf{x}(0))$ for all $t \ge 0$.  The bounds on $P$ further imply that $\|\mathbf{x}(t)\|^2 \le q^* \| \mathbf{x}(0)\|^2$. These LMI constraints are conservative in general and hence $q^*$ is a (possibly non-tight) upper bound on the MTEG.

\subsection{Global MTEG Analysis using Quadratic Constraints}
\label{sec:global_MTEG_Analysis}

An optimization problem similar to Eq.~\eqref{eq:linear_MTEG} can be formulated to study the MTEG in the nonlinear WKH system.  The lossless property for the nonlinear term in Eq.~\eqref{eq:lossless_constraint} can again be used as a global constraint.
\msh{Taking a similar approach as in Section~\ref{sec:QC_lossless}, we perturb the Lyapunov inequality to ensure a feasible solution can exist when only the lossless constraint is used.  This yields the following optimization for a given $\epsilon>0$:}

\begin{align}
\begin{split}
q^* :=\min  \quad & q\\
\mbox{ subject to:}~&  \text{I}\leq P \leq q\text{I}~, \\
&  \begin{bmatrix} A^TP + PA & P\\
P & \mathbf{0}\end{bmatrix} + \xi_{p_0} M_0 + \begin{bmatrix} \epsilon P & \mathbf{0}\\
    \mathbf{0} & \mathbf{0} \end{bmatrix} \le 0
\end{split}
\label{eq:Global_MTEGcalc_nonlinear}
\end{align}
Equation~\eqref{eq:Global_MTEGcalc_nonlinear} is now a SDP in the variables $(P,q,\xi_{p_0})$. As before, the LMI constraints imply that $V(\mathbf{x}):= \mathbf{x}^T P \mathbf{x}$ is a Lyapunov function for the system so that $V(\mathbf{x}(t)) \le V(\mathbf{x}(0))$ for all $t \ge 0$. The bounds on $P$ further imply that $\|\mathbf{x}(t)\|^2 \le q^* \| \mathbf{x}(0)\|^2$.

\subsection{Local MTEG Analysis using Quadratic Constraints}
\label{sec:local_MTEG_Analysis}
The ability to obtain MTEG bounds is of interest even beyond the globally stable regime. Here, we use the local properties of the nonlinearity derived in Section~\ref{sec:local_stab_analysis} to study the ``local'' MTEG performance in the nonlinear system for $Re>20$.

A formulation similar to Eq.~\eqref{eq:Global_MTEGcalc_nonlinear} can be used to study the effect of nonlinearity on MTEG in the nonlinear system. To perform the local MTEG analysis, additional local constraints are added to the optimization problem listed in Eq.~\eqref{eq:Global_MTEGcalc_nonlinear}. The local constraints that capture input-output properties of the nonlinear term are captured by $M_i$ for $i=1$ to $4$. The addition of these constraints, now facilitates the study of local MTEG on the nonlinear system. The local MTEG for the nonlinear system is computed via the following convex optimization:

\begin{align}
\begin{split}
\min  \quad & q\\
\mbox{ subject to:}~&  \text{I}\leq P \leq q\text{I}~, \\
&  \xi_{p_i} \geq 0 \qquad (\text{for}~i=1~\text{to}~4)~,\\
&  \begin{bmatrix} A^TP + PA & P\\
P & \mathbf{0}\end{bmatrix} + \xi_{p_0}M_0 + \sum_{i=1}^4 \xi_{p_i}M_i <0.
\end{split}
\label{eq:Local_MTEGcalc_nonlinear}
\end{align}
We will identify MTEG for the system about a local equilibrium point $\mathbf{x}_e=\mathbf{0}$ by solving this optimization for $P,~q$,~$\xi_{p_0} ,~\text{and}~\xi_{p_i}$ (for $i = 1~\text{to}~4$). The proof in Appendix~\ref{sec:appendixA} also applies for Eq.~\eqref{eq:Local_MTEGcalc_nonlinear}, therefore ensuring the states always remain inside the invariant set for all time $t\ge0$. 
\subsection{Results: MTEG analysis using quadratic constraints}

\subsubsection{Global MTEG analysis}
\msh{By solving the SDP in Eq.~\eqref{eq:Global_MTEGcalc_nonlinear} \ak{with $\epsilon=10^{-6}$}, we find that the MTEG bound is unity 
for all $Re<20$.}
\msh{It is interesting to note that the linear part of the WKH system exhibits unity MTEG for $Re<20$ as well.  Yet, the same MTEG bound from Eq.~\eqref{eq:Global_MTEGcalc_nonlinear} is stronger because it applies to the nonlinear system $F_u(L,N)$ with a lossless nonlinearity.}

\subsubsection{Local MTEG analysis}
For a given region $R$, we investigate the effect of the nonlinearity on MTEG by solving the convex optimization problem in Eq.~\eqref{eq:Local_MTEGcalc_nonlinear} for various $Re$. FIG.~\ref{fig:local_stability} indicates that the MTEG for the nonlinear system converges to that of the linear system for small values of $R\approx 10^{-3}$ and less.
However, when the size of $R$ is increased to $R = 10^{-2}$, we observe that the nonlinear terms exacerbate the MTEG relative to the linear case for $Re \geq 160$. Similarly, further increasing the region size to $R = 10^{-1}$ shows that for $Re >50$, the MTEG values are higher than those observed in the case with $R=10^{-2}$. This further shows that as nonlinear effects are increased, the MTEG of the system increases. Also, a significant increase in MTEG shows up at a much lower $Re$ by increasing $R$. It should be noted that  for the case with $R=10^{-1}$, the convex optimization problem becomes infeasible beyond $Re=70$. From this analysis, we note that the nonlinear term in the WKH model cannot be ignored, as it has significant influence on MTEG of the system. This is consistent with findings reported by Kerswell in \citep{Kerswell2018}, that the nonlinear NSE can exhibit a larger growth in energy compared to the linearized NSE.

\begin{figure}[!htb]
    \centering
    \includegraphics[scale=0.6,trim= 0.1cm 2.8cm 0.1cm 0.1cm]{Figs_New_V1/ReVsMTEG_VariousR_FontSize55_v4.eps}
    \caption{As $R$ and $Re$ increase, nonlinear effects become more substantial and MTEG increases.}
    \label{fig:ReVMTEG}
\end{figure}

\section{Lagrange Multiplier Analysis: Drawing Physical Insights Into Nonlinear Flow Interactions}
\label{sec:Lagrange_Mul_Analysis}

In addition to providing a framework to analyze stability and transient energy growth, the quadratic-constraints-based methods can be used to gain insights into the physics and dominating mechanisms underlying these dynamics. To do so, we analyze the Lagrange multipliers obtained after solving the convex optimization problem for a given $Re$ and local region~$R$. The Lagrange multipliers provide information on how the objective function is changing with respect to the constraints, which highlights the importance of the corresponding constraints in the optimization problem. 

  Dominant Lagrange multipliers can be identified by plotting their values over various $Re$.
  In FIG.~\ref{fig:Lagrange_multiplier}, we show the Lagrange multipliers obtained during the MTEG analysis of the WKH system for $R=0.01$. Each Lagrange multiplier is associated with their corresponding nonlinear terms. Here, we see that $\xi_{p_2}$---associated with the nonlinear term $\delta w^2$---is approximately 100 times more dominant than $\xi_{p_3}$---associated with the nonlinear term $\gamma w u -\delta w v$. Further, $\xi_{p3}$ is orders of magnitude larger than the multipliers associated with the other nonlinearities. Similar trends are observed for other values of $R$. This points to the fact that the nonlinear terms $\delta w^2$ and $\gamma w u -\delta w v$ are the dominant flow interactions contributing to MTEG in the WKH system.
  We obtain similar findings related to dominating flow interactions when comparing Lagrange multipliers obtained from the stability analysis results as well. Waleffe discusses the importance of the nonlinearities $\delta w^2$ and $\gamma w u -\delta w v$ in feeding $\dot{v}$ and $\dot{w}$, thereby serving central roles in sustaining turbulence and conserving energy, respectively. 
We note that this analysis of Lagrange multipliers allowed the same dominant nonlinear flow physics to be identified without reliance upon any prior knowledge or physical insight.
\begin{figure}[h!] 
    \centering
    \includegraphics[scale=0.65,trim = 1cm 1.5cm 1cm 1cm 1cm]{Figs_New_V1/ReVPenaltyFuncs_R_0_01_FontSize55.eps}
    \caption{The Lagrange multipliers shown against various $Re$. The two dominating nonlinear terms can be identified by analyzing the dominant Lagrange multipliers for $R=0.01$.}
    \label{fig:Lagrange_multiplier}
  \end{figure}
 
To demonstrate the dominance of these nonlinear interactions, we perform MTEG analysis while retaining only the constraints associated with the dominating nonlinear interactions and neglecting the other local interactions. 
We choose $R=0.01$ as before, but now use only the lossless constraint along with constraints associated with $\xi_{p_2}$ and $\xi_{p_3}$ (see green line in FIG.~\ref{fig:3_constraints}) and compare results with the case where all the constraints are retained (see black line in FIG.~\ref{fig:3_constraints}). 
The MTEG profile for system with both dominating nonlinear interactions~($\delta w^2,~\gamma w u -\delta w v$) closely approximates the response of the whole nonlinear system (shown by the black line in FIG.~\ref{fig:3_constraints}).
We observe similar qualitative trends for any other value of $R$ for which the optimization problem is feasible.

  \begin{figure}[!htb] 
     \centering
     \includegraphics[scale=0.55,trim = 1cm 1.5cm 1cm 1cm 1cm]{Figs_New_V1/ReVsMTEG_R_0_01_DominatingConstraints_FontSize55.eps}
     \caption{Local MTEG analysis with global lossless constraint and two most dominant constraints compared against the MTEG of linear system as well as the system with global and all local constraints for $R=0.01$.}
     \label{fig:3_constraints}
 \end{figure}

\section{Conclusion}
\label{sec:Conclusion}
In this work, we presented a quadratic constraints framework to perform stability and transient energy growth analysis of nonlinear systems. The proposed framework facilitates stability and transient energy growth analysis in global and local settings around a given equilibrium point.
The framework uses exact information from the linear dynamics,
while nonlinear interactions are replaced by quadratic constraints that capture input-output properties of the nonlinearity.
%

We demonstrated the proposed analysis approach on the WKH 
model of transitional and turbulent flow.
We first study the stability of the WKH model, for which the linear part is globally asymptotically stable for all $Re$.
It is found that the nonlinear WKH system is not globally stable for  $Re> 20$, consistent with previous results found in the literature. It is also observed that the energy conserving nonlinear terms eventually destabilize the system beyond the globally stable regime.
We also introduced a method for conducting maximum transient energy growth analysis when the system is globally asymptotically stable.
It was found that the maximum transient energy growth was
unity below the critical Reynolds number for global stability.

In order to assess stability and maximum transient energy growth
performance beyond the globally stable regime, we introduced
a new ``local'' analysis framework to analyze the system's local stability and transient energy growth properties.
Using this local analysis framework, we found that the nonlinear terms destabilize the system and also increase the maximum transient energy growth relative to the linear system.
Lastly, analyzing the Lagrange multipliers associated with each local constraint provided further insights into the physics.
By comparing the relative magnitudes of the Lagrange multipliers,
we are able to identify the dominating nonlinear interactions in the system, without any a priori knowledge of the flow physics.
The dominant nonlinear terms identified by this analysis
are in agreement with the physical mechanisms originally described
in~\cite{WaleffeSIAM1995}.
%

The proposed quadratic constraints framework for nonlinear stability analysis extends linear analysis techniques to account for nonlinear interactions via quadratic constraints. 
This enables a low-complexity framework for conducting global and local stability, input-output, and transient energy growth analyses of nonlinear flows. By replacing the nonlinearity with quadratic constraints greatly simplifies the ensuing analyses.
In addition, the use of quadratic constraints created an ability to identify dominant nonlinear interactions and to extract physical insights about underlying mechanisms underlying the complex dynamics of flow instabilities. 
\section{Acknowledgements}
This material is based upon the work supported by the Air Force Office of Scientific Research under award number FA 9550-19-1-0034, monitored by Dr. Gregg Abate, and the National Science Foundation under award number 1943988, monitored by Dr. Ronald D. Joslin.

\bibliographystyle{aiaa}
\bibliography{IQC-WKH-PRF20}

\clearpage
\appendix

\section{ The set bounded by local region $R$ is invariant}
\label{sec:appendixA}
The role of $R$ in this analysis can be made more precise. Assume there is a feasible solution $P>0$ for the linear matrix inequality in Eq.~\eqref{eq:Local_Stab_LMI}. Then the Lyapunov function $V(\mathbf{x})=\mathbf{x}^TP\mathbf{x}$ satisfies $dV(\mathbf{x}(t))/dt <0 $ as long as $\mathbf{x}(t)^T\mathbf{x}(t) \le R^2$.  This implies that trajectories converge back to $\mathbf{x}_e=0$ if the initial conditions are sufficiently close to the origin. In particular, the constraint $P>I$ implies that $\mathbf{x}^T\mathbf{x}< V(\mathbf{x})$. A simple proof by contradiction can be used to demonstrate that if $V(\mathbf{x}(0))<R^2$ then: (i) the trajectory $\mathbf{x}(t)$ remains in the local region $\|\mathbf{x}(t)\|^2\le R^2$ and (ii) the trajectory $\mathbf{x}(t)$ decays to the origin.  In summary, the set $S_R := \{ \mathbf{x}\,:\,V(\mathbf{x})<R^2\}$ is a domain of attraction.\\

\underline{\textbf{Proof:}} \\
Define the set $S_R:=\{\mathbf{x}\,:\,V(x)<R^2\}$. Assume $\mathbf{x}(0) \in S_R$ and
let $\mathbf{x}(t)$ denote the corresponding state trajectory from this initial condition.  Assume there exists a time $T_1$ such that  $\mathbf{x}(T_1) \notin S_R$ and let $T_0$ be the smallest (infimum) of times such that $\mathbf{x}(t) \notin S_R$. The solution $\mathbf{x}(t)$ is a continuous function of time and hence $\mathbf{x}(t) \in S_R$ for all $t \in [0,T_0)$ and, moreover, $\mathbf{x}(t)$ is on the boundary of $S_R$ so that $V(\mathbf{x}(T_0))=R^2$. Therefore, the local quadratic constraints are valid for all $t\in [0,T_0]$.  As noted above, $P>I$ implies that if $\mathbf{x}(t) \in S_R$ then $\|\mathbf{x}(t)\|^2 <R^2$. The constraints in Eq.~\eqref{eq:Local_Stab_LMI} imply that, for a sufficiently small $\epsilon>0$, the Lyapunov function satisfies $dV(\mathbf{x}(t))/dt \le -\epsilon \mathbf{x}(t)^T \mathbf{x}(t) \ \forall \ t\in [0,T_0]$.
Integrating yields the following bound for any $\mathbf{x}(0) \ne 0$:

\begin{align}
    V(\mathbf{x}(T_0)) \le V(\mathbf{x}(0)) < R^2
\end{align}
 This contradicts the assumption that $V(\mathbf{x}(T_0))=R^2$ and hence trajectories must remain in $S_R$. Moreover, the Lyapunov condition $dV/dt \le -\epsilon \mathbf{x}(t)^T\mathbf{x}(t) \ \forall \ t\in [0,T_0]$ implies that the trajectories in this region decay asymptotically back to the origin.


\section{Sector-bounded nonlinearities}
\label{sec:appendixB}
\begin{figure}[!h]
    \centering
    \includegraphics[scale=0.5]{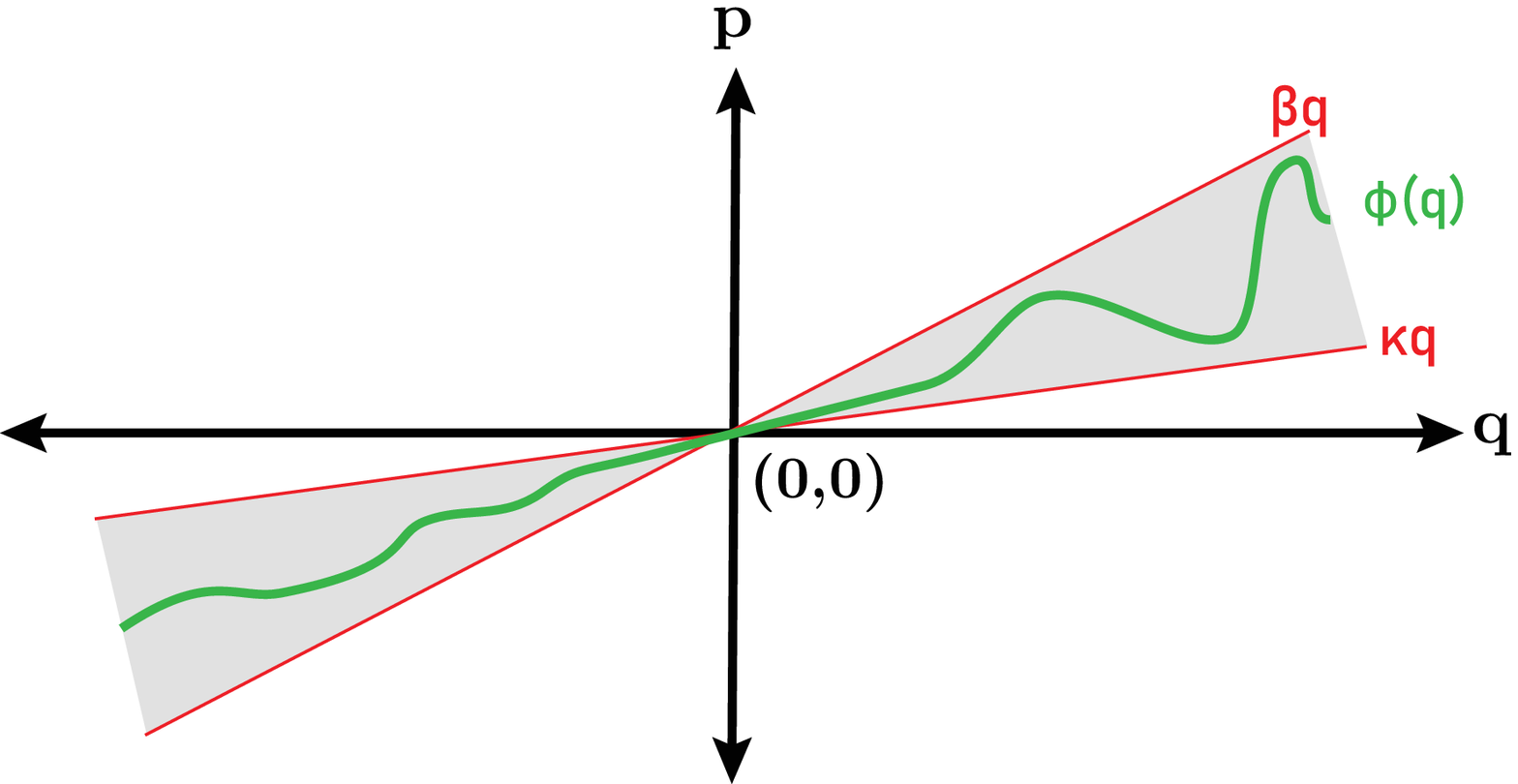}
    \caption{Illustration of a generic sector bounded nonlinearity}
    \label{fig:sector_nonlinearity}
\end{figure}

A comprehensive review of sector bounded systems can be obtained in~\citep{Willems1972,Khalil}. Given a nonlinearity $\phi : \R \rightarrow \R$, $\phi$ lies in a sector $[\kappa,\beta]$ if for all $q \in \R$, $p=\phi(q)$ 
lies between the lines of slope $\kappa$ and $\beta$ at each point in time. This property can also written in terms of the input and output of the nonlinearity as a quadratic inequality of the form $(\beta q-\phi(q))(\phi(q)-\kappa q) \ge 0~\forall q\in\R$, or equivalently

\begin{align}
 \begin{bmatrix}q \\ p\end{bmatrix}^T \begin{bmatrix} -\kappa \beta & \frac{1}{2}(\kappa + \beta) \\ \frac{1}{2}(\kappa + \beta) & -1 \end{bmatrix}\begin{bmatrix}q \\ p\end{bmatrix} \ge 0 ~\forall q\in\R.
 \label{eq:appendix_scalar_sector_constraints}
\end{align}
Graphically this is shown in FIG.~\ref{fig:sector_nonlinearity}, where the shaded region contains the nonlinearity $\phi$.

\end{document}